\documentclass[apj]{emulateapj}
\usepackage{natbib}
\usepackage{apjfonts}
\usepackage{graphics}
\newcommand{\hb}{H$\beta$}
\newcommand{\ha}{H$\alpha$}
\newcommand{\mbh}{M$_{\rm BH}$}
\newcommand{\msigma}{M$_{\rm BH}-\sigma_{*}$}

\newcommand{\ergs}{erg s$^{-1}$}
\newcommand{\LPAH}{L$_{\rm 3.3~PAH}$}

\begin{document}
\title{The connection between 3.3 \micron~PAH emission and AGN activity}
\author{Jong-Hak Woo$^{1}$, Ji Hoon Kim$^{2}$, Masatoshi Imanishi$^{3,4,5}$, and Dawoo Park$^{1}$}
\affil{$^{1}$Astronomy Program, Department of Physics and Astronomy,
Seoul National University, Seoul, Republic of Korea\\
$^{2}$ Center for the Exploration of the Origin of the Universe,
Astronomy Program, Department of Physics and Astronomy,\\
Seoul National University, Seoul, Republic of Korea\\
$^{3}$ Subaru Telescope, 650 North A'ohoku Place, Hilo, Hawaii, 96720,
U.S.A.\\
$^{4}$ National Astronomical Observatory of Japan, 2-21-1
Osawa, Mitaka, Tokyo 181-8588, Japan\\
$^{5}$ Department of Astronomy, School of Science, Graduate
University for Advanced Studies (SOKENDAI), Mitaka, Tokyo 181-8588,
Japan\\
}
\email{woo@astro.snu.ac.kr}

\begin{abstract}

We investigate the connection between starburst and active galactic nucleus (AGN) activity 
by comparing the 3.3 \micron~polycyclic aromatic hydrocarbon (PAH) emission with AGN properties.
Utilizing the slit-less spectroscopic capability of the AKARI space telescope, 
we observe moderate-luminosity Type I AGN at z$\sim$0.4 to measure global
starburst activity. 
The 3.3 \micron~PAH emissions are detected for 7 out of 26 target galaxies. 
We find no strong correlation between the 3.3 \micron~PAH emission and AGN luminosity
in the limited range of the observed AGN luminosity,
suggesting that global star formation may not be closely
related to AGN activity. 
Combining our measurements with the previous 3.3 \micron~measurements
of low redshift Type I AGN in the literature, 
we investigate the connection between nuclear starburst
and AGN activity. In contrast to global star formation, the 3.3 \micron~PAH 
luminosity measured from the central part of galaxies correlates with AGN luminosity,
implying that starburst and AGN activity are directly connected in the nuclear
region. 
\end{abstract}


\section{Introduction}
\label{intro}

The correlations of black hole mass with galaxy properties 
imply a close link between black hole growth and galaxy evolution 
\citep{FM00, Gt00, gultekin+09, woo10}.
Numerous theoretical and observational studies have been dedicated 
to revealing the nature of the connection
(e.g., Kauffmann \& Haehnelt 2000; Springel et al. 2005; 
Croton et al. 2006; Woo et al. 2006, 2008; Hopkins et al. 2007; 
Merloni et al. 2010; Bennert et al. 2011).

Probing the direct connection between starburst and  
active galactic nuclei (AGN) can shed light on how black hole
growth and bulge growth are inter-connected. 
However, it is challenging to observationally measure both the starburst rate 
and AGN activity for given galaxies. 
Optically-identified broad-line AGNs are the best objects for measuring
black hole mass and AGN accretion rate. On the other hand the star formation rate of
host galaxies is difficult to measure 
since AGN flux dominates in the UV continuum and hydrogen recombination lines,
i.e., \ha, which are generally used as a star formation indicator 
in non-AGN galaxies.
In the case of Type II AGN, it is relatively easy to measure star formation
rate while black hole mass and accretion rate are difficult to determine
owing to the dust obscuration of accretion disk and broad-line region.

The polycyclic aromatic hydrocarbon (PAH) emission features emitted by small 
carbon grains are one of the star formation indicators.
Thus, PAH emissions have been used to constrain star formation 
activity in the host galaxies of Type 1 AGN (e.g., Ogle et al. 2006; Schweitzer et al. 2006;
Shi et al. 2007).
Based on the ground-based and space observations, various PAH emission features,
i.e., 3.3, 6.2, 7.7, 8.6, 11.3, and 12.7 \micron~lines,
have been detected in many local Seyfert 1 galaxies and distant QSOs 
although not all AGN show PAH emission features  
\citep[e.g.,][]
{clavel+00, Rod03, imanishi04, ntz07,lz07,lutz+08,shi+09,imanishi10,Oi10,imanishi11}. 

Among PAH-detected AGN, there seems to be a correlation between
nuclear starburst and AGN activity. Using
a sample of low-redshift Palomar-Green (PG) QSOs observed with 
the Spitzer Infrared Spectrograph (IRS), \citet{ntz07} 
showed that 7.7 \micron~PAH luminosity increases 
with AGN continuum luminosity at 5100\AA~ (L$_{\rm 5100\AA}$), indicating the starburst-AGN connection.
High-redshift QSOs seem to follow the same trend, extending the correlation
to much higher luminosity range \citep{lutz+08}.
The starburst-AGN connection is supported by a number of theoretical studies,
demonstrating that both nuclear starburst events 
and AGN activity
can be triggered by gas inflow through galaxy interactions
(e.g., Kauffmann \& Haehnelt 2000; Robertson et al. 2006; 
Ciotti \& Ostriker 2007; Hopkins et al. 2009).

For lower luminosity AGNs it is less clear how AGN activity 
is connected to starburst since AGN power is more difficult to measure
owing to increasing host galaxy contribution in lower luminosity AGN.
Based on the ground-based L-band slit spectra, 
\citet{imanishi04} measured the 3.3 \micron~PAH luminosity of local Seyfert 1 and 2 galaxies,
showing a positive correlation between starburst and 
AGN activity although the correlation is not tight with a large scatter.
In their study, radio luminosity and N-band luminosity within 1.5''
of the host galaxies are used as an indicator of AGN power.
Using 2-10 kev X-ray luminosity, \citet{Wt08} showed a similar trend 
between AGN power and the 3.3 \micron~PAH luminosity for a larger sample
of local Seyfert 1 and 2 galaxies. 

Most previous observational studies on PAH features with ground based 
long-slit spectroscopy or Spitzer IRS observations have probed the 
inner part of galaxies, indicating a connection between {\it nuclear} starburst within a few kpc
and AGN activity \citep[e.g.,][]{Wt08}. 
In contrast, whether global star formation over
entire galaxies is related with AGN activity it is not well studied 
since most previous studies, particularly for local and low-redshift galaxies,
have used a relatively small aperture to extract PAH emission features, 
measuring nuclear PAH emissions.
Based on the spatially resolved mid-IR spectroscopy of nearby active galaxies, 
several studies have reported that 
PAH emissions are mainly from the disk \citep[e.g.,][]{laurent+00, lefloch+01},
suggesting that the global PAH luminosity measured over entire galaxies
may be much larger than nuclear PAH luminosity.

In this paper, we investigate global and nuclear star formation in the
host galaxies of Type 1 AGN, using the 3.3 \micron~PAH emission measurements 
based on our new observations with the AKARI space telescope and 
collected data from the literature.
The AKARI telescope provides a unique capability of slit-less spectroscopy
to measure global star formation over entire galaxies     
(e.g., Imanishi et al. 2010).
This paper is organized as follows. In \S \ref{data}, we describe 
sample selection, 
observations and data reduction of the AKARI observations. 
We present 3.3 \micron~PAH emission measurements in \S \ref{analysis}. 
Main results on the starburst-AGN connection are presented in \S \ref{results}.
Discussion and conclusions are followed in \S~5.
Throughout the paper, we assume a Hubble constant of
H$_{o}$ = 70 km s$^{-1}$ Mpc$^{-1}$, $\Omega_{\Lambda} =0.7$, and $\Omega_{M} = 0.3$.

\section{Observations and Data Reduction}
\label{data}

\subsection{Sample Selection}
\label{sample}

We selected a sample of 27 moderate-luminosity AGN (L$_{\rm 5100\AA}$ $\sim$ 0.1-3 $\times$ 
10$^{44}$ \ergs) at 0.35 $<$ z $<$ 0.36 from the Sloan Digital 
Sky Survey Data Release 7 to study star formation activity based on the PAH feature
at 3.3 \micron. The sample was initially selected for investigating the evolution 
of the black hole mass-stellar velocity dispersion (\msigma) relation. Details of sample selection can be found in \citet{woo06} and \citet{treu07}. 
In summary, we selected broad-line AGNs with the H$\beta$ equivalent width (EW) $>$ 5 \AA~ 
to properly estimate black hole masses based on the kinematics of broad-line region gas. 
The specific redshift range was selected to avoid
sky emission lines on the stellar absorption features, from which stellar velocity dispersions
were directly measured. 
The spectral energy distribution of 24 objects in the sample was in detail investigated
by \citep{szathmary11} 
based on the multi-wavelength data including Chandra X-ray observations, HST imaging, 
and Spitzer IRAC and MIPS imaging as well as the archival data of GALEX and 
Two Micron All-Sky Survey (2MASS).
From the spectral energy distribution (SED) fitting analysis, 
AGN SED was separated from galaxy SED, 
leading to accurate measurements of AGN bolometric luminosities.
Black hole mass of each AGN was estimated by the single-epoch virial method,
which combines the line width of \hb~and the optical luminosity at 5100\AA (L$_{5100}$), 
utilizing the size-luminosity relation of the reverberation sample (Bentz et al.
2009) and the most recent calibration of the virial factor by \citet{woo10} (see Table 1). 
The black hole mass of the sample covers an order of magnitude, i.e., $7.6 < \log$ \mbh/M$_{\odot}$ $< 8.7$.
Combining black hole mass estimates with the bolometric luminosities measured   
from the SED fitting, we obtained the Eddington ratios, ranging from  a few to $\sim$50\%
of the Eddington limit. The basic properties of the sample are listed in Table \ref{table1}.

\begin{deluxetable*}{cccccccccc}
\tabletypesize{\scriptsize}
\tablecaption{Properties of the AKARI sample: Type I AGN at z$\sim$0.4
\label{table1}
}
\tablewidth{0pt}
\tablehead{
\colhead{Name} 
&\colhead{RA (J2000)} &\colhead{DEC (J2000)}
&\colhead{redshift}
&\colhead{log L$_{bol}$}
&\colhead{$\lambda$L$_{5100,nuc}$}
&\colhead{$f_{nuc}$}
&\colhead{log L$_{X}$}
&\colhead{log M$_{\rm BH}$}
&\colhead{pointings} \\
\colhead{} 
&\colhead{} &\colhead{}
&\colhead{}
&\colhead{(erg/s)}
&\colhead{(erg/s)}
&\colhead{}
&\colhead{(erg/s)}
&\colhead{(M$_{\odot}$)}
&\colhead{}    \\
\colhead{(1)}  &
\colhead{(2)}  &
\colhead{(3)}  &
\colhead{(4)}  &
\colhead{(5)}  &
\colhead{(6)}  &
\colhead{(7)}  &
\colhead{(8)}  &
\colhead{(9)}  &
\colhead{(10)} \\
}
\startdata

S01	& 15 39 16.23 & +03 23 22.1 & 0.3592	& 45.23 & 0.72 & 0.29 & 44.48 & 8.20 	 &1\\
S02	& 16 11 11.67 & +51 31 31.1 & 0.3544	& 45.13 & 0.34 & 0.22 &	44.60 & 7.98  &  4	\\
 S03	& 17 32 03.11 & +61 17 52.0  & 0.3583	& 45.50 & 1.64 & 0.39 & 44.95 & 8.28  & 4\\
 S04	& 21 02 11.51 & -06 46 45.0 & 0.3578	& 45.27 & 1.33 & 0.36 &	44.55 & 8.44  &4 \\
 S05	& 21 04 51.85 & -07 12 09.4 & 0.3530  	& 45.44 & 1.85 & 0.47 & 44.84 & 8.74 &4\\
 S06	& 21 20 34.19 & -06 41 22.2 & 0.3684  	& 45.05 & 0.51 & 0.18 &	44.25 & 8.16 &4\\             
 S07	& 23 09 46.14 & +00 00 48.9 & 0.3518  	& 45.47 & 2.10 & 0.45 &	44.80 & 8.53  &2\\ 
 S08	& 23 59 53.44 & -09 36 55.5 & 0.3585	& 45.23 & 1.22 & 0.40 & 44.55 & 8.10    &4\\
 S09	& 00 59 16.11 & +15 38 16.1  & 0.3542	& 45.37 & 0.71 & 0.22 & 44.78 & 8.13  &4\\
S10	& 01 01 12.07 & -09 45 00.8  & 0.3506	& 45.48 & 1.02 & 0.27 & 44.93 & 8.25 &4\\
 S11	& 01 07 15.97 & -08 34 29.4 & 0.3557	& 45.34 & 0.52 & 0.14 & 44.85 & 8.00	  &4\\ 
 S12	& 02 13 40.60 & +13 47 43.3  & 0.3570	& 45.43 & 0.97 & 0.28 & 44.62 & 8.67	  &4\\
 S21	& 11 05 56.18 & +03 12 43.3  & 0.3534  	& 45.88 & 2.15 & 0.34 &	44.06 & 8.79   &1\\
 S23	& 14 00 16.66 & -01 08 22.2  & 0.3510  	& 45.42 & 1.11 & 0.29 & 44.75 & 8.70  &3\\ 
 S26	& 15 29 22.26 & +59 28 54.6  & 0.3691	& 45.13 & 0.52 & 0.27 & 44.38 & 8.02	   &4\\
 S27	& 15 36 51.28 & +54 14 42.7 & 0.3667	& 45.23 & 0.95 & 0.36 & 44.74 & 8.10  &4\\
 S28	& 16 11 56.30 & +45 16 11.0 & 0.3680  	& 45.13 & 0.76 & 0.03 & 44.66 & 7.90  &4\\ 
 S29	& 21 58 41.93 & -01 15 00.3 & 0.3575  	& 45.43 & 0.59 & 0.10 & 44.54 &	7.94  &4\\
 S31	& 10 15 27.26 & +62 59 11.5 & 0.3504	& 45.38 & 0.29 & 0.08 & 44.58 & 7.94  &2\\
 SS1	& 08 04 27.99 & +52 23 06.2 &	0.3566	& 45.41 & 0.39 & 0.11 & 44.89 & 7.75	 &4\\
 SS2	& 09 34 55.60 & +05 14 09.1 & 	0.3675	& 45.21 & 0.33 & 0.13 & 44.54 & 7.72  & 	3\\
 SS9	& 12 58 38.71 & +45 55 15.5 & 0.3704	& 44.92$^a$ & 0.93 &0.26& \nodata & 8.05	&4\\
 SS10   & 13 34 14.84 & +11 42 21.5 & 0.3658	& 45.30$^a$ & 2.26 &0.43& \nodata & 7.94   &4\\
 SS11   & 13 52 26.90 & +39 24 26.8 & 0.3732	& 44.66$^a$ & 0.51 &0.15& \nodata & 8.11 &4\\
 SS12   & 15 01 16.83 & +53 31 02.4 & 0.3626	& 45.46$^a$ & 3.24 &0.52& \nodata & 8.15 &4\\
 SS13   & 15 05 41.79 & +49 35 20.0 & 0.3745    & 44.94$^a$ & 0.98 &0.38& \nodata & 7.63  &4
\enddata

\tablecomments{ 
Col. (1): Object name.
Col. (2): RA.
Col. (3): DEC.
Col. (4): redshift
Col. (5): AGN bolometric luminosity measured from multi-wavelength data,
adopted from \citet{szathmary11}. 
Col. (6): AGN monochromatic luminosity at 5100\AA~ in 10$^{44}$ \ergs~ units, 
corrected for host galaxy starlight based on the HST images, 
taken from \citet{treu07, bennert10}.
Col. (7): AGN fraction in the observed flux at 5100\AA, adopted from
\citet{treu07, bennert10}.
Col. (8): AGN X-ray luminosity from \citet{szathmary11}.
Col. (9): Black hole mass estimated from \hb~line width and L$_{5100}$ 
adopted from \citet{woo06,bennert10}.
Col. (10): Number of pointings for AKARI observation
}

\tablenotetext{a}{AGN bolometric luminosity is calculated from 
optical luminosity (L$_{5100}$) assuming that 
L$_{bol}$ = 9 $\times \lambda$ L$_{5100}$.}  

\end{deluxetable*}

\subsection{AKARI Observations}
\label{obs}

To detect the PAH features at the rest-frame $\sim$3.3 \micron, we carried out spectroscopic 
observations using the Infrared Camera (IRC) on board the AKARI space telescope
\citep{IRC}. The field of view of IRC is roughly 10' $\times$ 10' with 
a spatial scale of 1.45''/pixel.
The full-width-at-half-maximum of the point spread function is 3.2 pixels, 
corresponding to 4.64''.  
The NIR grism mode (NG) was adopted for most of the observations, 
providing spectral resolution of R $\sim$ 120 at 3.6 \micron~ for a point source \citep{IRC}.
A unique capability of IRC is the slit-less spectroscopic mode. 
With 1' by 1' aperture used for pointing mode, the photons from the
entire galaxies in our sample can be detected. 
Thus, the measured 3.3 \micron~PAH luminosities represent
the {\it global} star formation.

The exposure time for each AGN host galaxy was determined in order to 
achieve signal-to-noise ratio S/N $>$ 10 in the continuum. 
We assigned 4 pointings (6 minute exposure per pointing) for each target
based on the exposure time calculation using the K-band magnitudes
from the 2MASS archive. 
Owing to the complexity of scheduling of the sun-synchronous
polar orbit of the AKARI, a couple of objects were observed with less than 4 points.
In summary, we obtained a total of 93 pointings for 27 target galaxies
(see \ref{table1}).

\subsection{Data Reduction}
\label{dr}

We carried out data reduction using the IRC spectroscopy pipeline\footnote{ http://www.ir.isas.jaxa.jp/ASTRO-F/Observation/DataReduction/IRC/}.
The data reduction procedures with the IRC spectroscopy pipeline 
is described in detail by \citet{IRCpipe}.
In summary, the reduction steps include standard procedures: 
dark subtraction, linearity correction, flat-fielding,
background subtraction, extraction of 2-dimensional spectra,  
wavelength calibration, and flux calibration.
Before extracting 1-dimensional spectra, 
we applied additional cosmic ray removal using {\it L.A.Cosmic} \citep{laC} 
to remove hot pixels, which significantly increased in Phase 3 (post liquid-He warm
mission cooled by the onboard cryocooler).
To extract 1-dimensional spectra, we used five pixels along the spatial 
direction as recommended by \citet{IRCpipe}.
As a consistency check, we compared the spectra extracted from
larger aperture sizes and found no difference in the extracted spectra.
Finally, we combine the extracted 1-dimensional spectra from each pointing
while applying additional sigma-clipping. 
In order to improve signal-to-noise ratio, we apply three pixels binning 
along the wavelength direction. 
We were not able to extract a spectrum for one object (S10) due to the low 
quality of the observed spectral images. Thus, this target is excluded
for the following analysis.

\begin{figure}
\begin{center}
\includegraphics[scale=0.5]{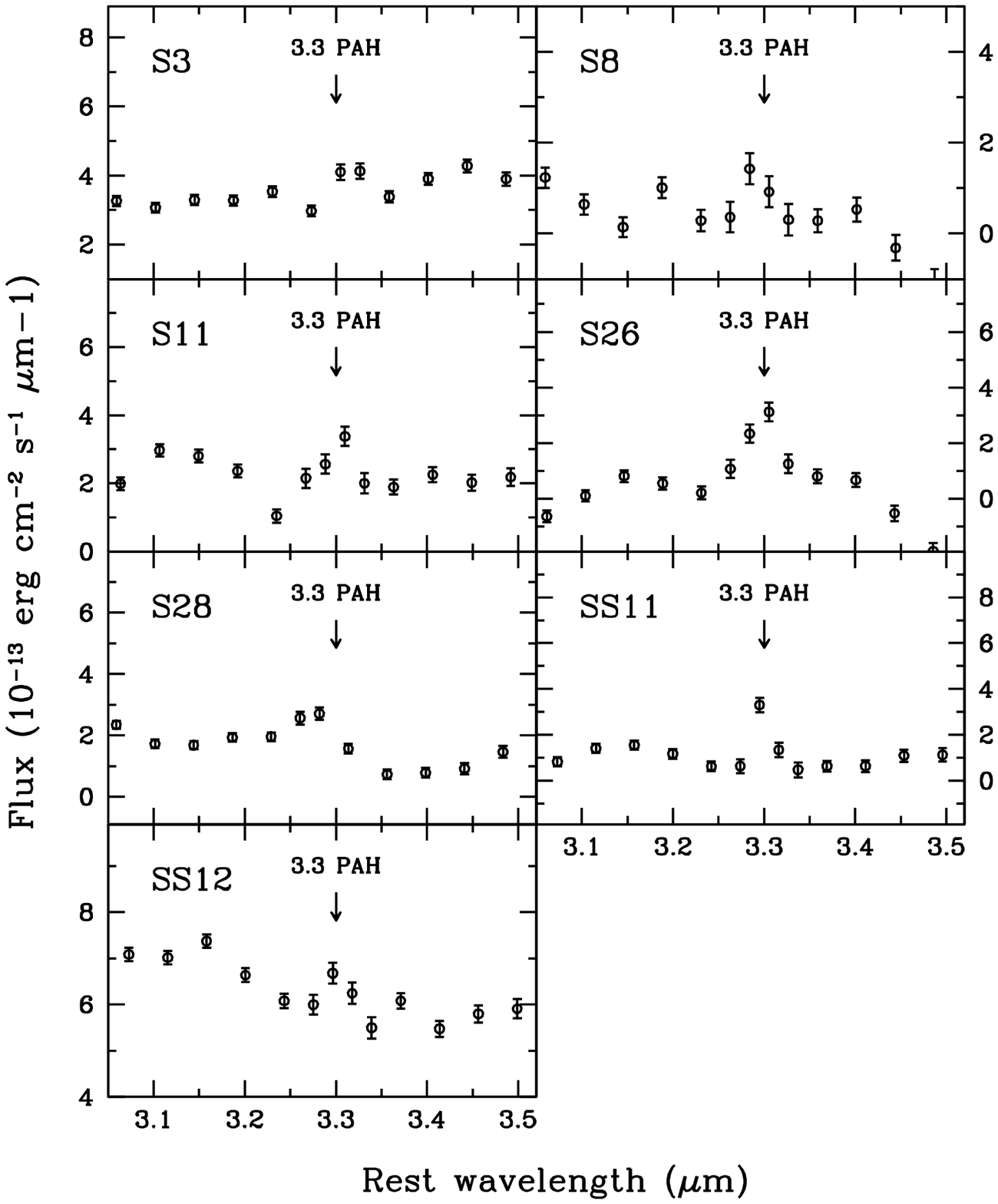}
\caption{AKARI spectra of moderate-luminosity AGN at z$\sim$0.4 with the 3.3 \micron~PAH emission detection. The location of the 3.3 \micron~PAH emission is indicated by an arrow in each 
panel. 
\label{fig1}}
\end{center}
\end{figure}

\section{3.3 \micron~PAH measurements}
\label{analysis}

\subsection{Individual galaxies}
\label{indiv}

In Figure \ref{fig1} we present the reduced spectra of 7 galaxies 
with the 3.3 \micron~PAH emission detection. 
Due to the low signal-to-noise ratio of the observed spectra,
it is necessary to assume an intrinsic profile of the 3.3 \micron~PAH emission line
in order to confirm the detection of the line and measure its strength.
In measuring the line strength of the PAH emission lines,
several different fitting schemes with various line profiles, 
i.g., Gaussian, Lorentzian, Drude profiles, and Spline functions have been devised. 
For example, \citet{Ga08} applied Spline functions and Lorentz profiles
for fitting PAH emission lines in various sources and found no
systematic difference in the measured PAH fluxes while
\citet{Sm07} applied Drude profile to mid-IR
spectra obtained from the $Spitzer$ IRS.
We chose a Gaussian profile, which is generally adopted for 
the 3.3 \micron~PAH emission lines observed in starburst galaxies
(Type-A sources in Tokunaga et al. 1991).
In practice, we applied a Gaussian profile with a fixed width
in the rest-frame for modeling the 3.3 \micron~PAH line.
We adopted 21 nm as a fixed width of the line profile, which has been
determined from the PAH line profile of M82 (Tokunaga et al. 1991) and 
used for our previous studies of the 3.3\micron~PAH line based on 
the AKARI IRC spectra (Imanishi et al. 2008, 2010).
 
Continuum subtraction is the largest source of uncertainty in measuring PAH
emission line flux since absorption features are abundant in the near- and mid-IR
spectral ranges. The vicinity of the 3.3 $\mu m$ PAH feature 
is not an exception: There is a plethora of absorption features in the rest-frame
2 - 5 \micron~range.
We applied a linear fit to the continuum in the 3.0 - 3.6 \micron~range
with two-pixel binning, 
after masking the location of the PAH feature around the 3.3 $\mu m$,
as similarly practiced in other studies \citep[e.g.][]{imanishi04,imanishi10,Oi10,imanishi11}. 

While fitting the 3.3 $\mu m$ PAH feature by a Gaussian profile,
we determined the peak and the central wavelength of the profile
based on the $\chi^2$ minimization. Then, the line flux and the equivalent width 
were determined from the fit (see Table \ref{table2}). 
The typical error of the measured flux density is $\sim 3 \times 10^{-14}$ erg s$^{-1}$ cm$^{-2}$ $\micron^{-1}$.

We detected the 3.3 \micron~PAH feature from 7  out of 26 
observed galaxies, resulting in a detection rate of 27\%. 
Compared to higher luminosity AGN, the 3.3 \micron~PAH detection rate of
our lower-luminosity AGN is slightly higher although the difference
is not easy to interpret owing to the different exposure time and flux limits. 
For example, \citet{imanishi11} 
investigated the PAH emission features of 30 PG QSOs based on the ground-based
observations, and detected the 3.3\micron~PAH emission from 5 QSOs,
resulting in a $\sim$ 17\% detection rate.
If we exclude S3 and SS12, for which 3.3\micron~PAH emission is marginally 
detected, then the detection rate of our lower-luminosity AGN decreases
to 19\%, suggesting that the detection rate is similar to that of PG QSOs.

Compared to the detection rate of other PAH lines at longer wavelength 
(e.g., 6.7, 7.7, 8.6, 11.2 \micron) in local Seyfert galaxies,
the detection rate of the 3.3\micron~PAH line in our sample seems to be much lower 
(see Sales et al. 2010). This difference may be due to the fact that
3.3\micron~PAH is much weaker than other lines at longer wavelength,
thus more difficult to be detected.
For better understanding of the relation among various PAH lines,
spectroscopic studies with a large spectral range including 3.3\micron~and
other PAH lines are required.

The detected 3.3 \micron~PAH line luminosity (\LPAH) ranges from $\sim 0.9-5.7 \times 10^{42}$ 
\ergs~ while the upper limit of the PAH luminosity of the undetected galaxies
is close to $\sim 10^{42}$ \ergs. 
Compared to the 3.3 \micron~PAH luminosity range
of Type I AGN host galaxies measured from ground-based observations 
(\LPAH=10$^{39}$ - 10$^{42}$ \ergs), 
the 3.3 \micron~PAH luminosities from the AKARI slit-less spectroscopy are more 
than a factor of 10 higher. This is probably due to the aperture effect, 
reflecting the difference between nuclear and global star formation. 
Ground-based observations generally used a narrow-slit to probe nuclear starburst 
while AKARI slit-less spectroscopy provides the global star formation
rate over entire galaxies. A direct comparison between ground-based
and AKARI observations supports this interpretation. For example,
a recent AKARI study by \citet{imanishi10} reported the 3.3 \micron~PAH
luminosity of NGC 7469 as $6.3\times 10^{41}$ \ergs~ while the ground-based
spectroscopy with a 1.6'' slit width by \citet{imanishi04} presented
an upper limit of $2.7\times 10^{40}$\ergs, indicating that the nuclear
PAH emission is a small fraction of the global PAH emission.

For galaxies with no 3.3 $\mu m$ PAH emission detection, we measured the upper limits.
First, we averaged the 1 $\sigma$ errors at the wavelength range of the 
expected 3.3 \micron~PAH line. Then, we assumed a Gaussian profile with a peak
flux, which is 3 times larger than the average 1 $\sigma$ error.
In the case of the line width of the profile, 
we used a constant width of 21 nm in the rest-frame.
The measurements and upper limits of the 3.3 \micron~PAH luminosity are listed
in Table 2.

\subsection{Combined Spectra}    
\label{stack}

\begin{figure}
\begin{center}
\includegraphics[angle=90, scale=0.4]{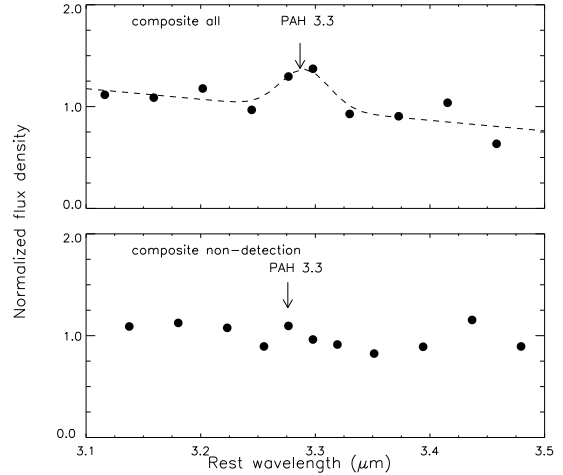}
\caption{Stacked spectra of the AKARI sample. The upper panel shows the spectrum stacked with the entire sample of 26 galaxies, while the bottom panel shows the spectrum spectrum stacked with the galaxies without the 3.3 \micron~PAH emission detected. These stacked spectra are the sum of normalized individual spectra within the wavelength range of 3.1 to 3.5 \micron~in the rest frame. We also apply two pixel binning to the data points outside the 3.3 \micron~PAH emission feature.
\label{fig2}}
\end{center}
\end{figure}

In order to construct the representative spectra of the sample
and to recover the undetected 3.3 \micron~PAH in individual galaxies, 
we constructed two combined spectra respectively for the entire sample and the galaxies 
with no 3.3 \micron~PAH emission detection. 
To construct stacked spectra, we used a wavelength range between 3.1 - 3.5 \micron~
in the rest frame and normalized individual spectra by a continuum fit.
Then, we summed individual spectra without weighting.
We also applied two-pixel binning to the data points outside of the 3.3 \micron~PAH line
to smooth the continuum. 
The stacked spectra of the entire sample and the galaxies with no 3.3 \micron~PAH detection
are presented in Figure \ref{fig2}. 

The combined spectrum of the entire sample clearly shows the 3.3 $\mu m$ PAH emission.
By fitting the PAH emission using a Gaussian profile with a width of 21 nm, 
we measured the EW of the 3.3 $\mu m$ PAH line. 
The EW of PAH in our sample is larger than that of QSOs. For example, 
based on a sample of PG QSOs, \citet{imanishi11} reported that
the EWs of the 3.3 \micron~PAH emission of individual galaxies are typically less than 10 nm
while the EW of the 3.3 \micron~PAH emission measured from their combined spectrum is only 1 nm.
The smaller EW of PAH emission in higher luminosity AGN may be
interpreted to mean that PAH lines are diluted by higher AGN continuum.
However, a larger sample is required to reveal the origin of the trend.

\begin{deluxetable}{cccc}
\tablecaption{Global 3.3 \micron~PAH emission measurements based on the AKARI observations
\label{table2} }
\tablewidth{0pt}
\tablehead{
\colhead{Object} 
&\colhead{f$_{3.3 \mu m}$}
&\colhead{L$_{3.3 \mu m}$} 
&\colhead{EW$_{3.3 \mu m}$}  \\
\colhead{} 
&\colhead{(10$^{-15}$ erg s$^{-1}$ cm$^{-2}$)}
&\colhead{(10$^{42}$ erg s$^{-1}$)} 
&\colhead{(nm)}   \\
\colhead{(1)}  &
\colhead{(2)}  &
\colhead{(3)}  &
\colhead{(4)}  \\
}

\startdata
 S01	 & < 5.09  & <2.19  & ....\\
 S02	 & < 10.47 & <4.36 & .... \\
 S03	 & 3.10    & 1.33  & 9\\
 S04	 & < 1.55& <0.66  & .... \\
 S05	 & < 2.00& <0.83  & .... \\
 S06	 &< 2.03& <0.93  & .... \\             
 S07	 &< 2.76 & <1.13  & .... \\ 
 S08	 & 3.80    & 1.63  & 87 \\
 S09     & < 31.41 & < 13.07 & .... \\
S10	& .... & .... & .... \\
 S11	&4.40& 1.85  & 21\\ 
 S12     & < 1.77  & < 0.75 & .... \\
 S21	 & < 12.73 & <5.27  & .... \\
 S23	 &<  1.97 &<0.80 & .... \\ 
 S26	 & 12.50 &5.73  & 206 \\
 S27	 & < 1.69  &<0.76 & .... \\
 S28	 &6.65 & 3.03  &43 \\ 
 S29	 & < 1.51 & <0.64 & .... \\
 S31	 & < 1.43  & <0.58 & .... \\
 SS1	 & < 1.77  & <0.74 & .... \\
 SS2	 & < 1.84  & <0.83 & .... \\
 SS3 & < 1.57 &<0.66 & .... \\
 SS9 & < 1.72 &<0.79 & .... \\
 SS10  & < 1.56 &<0.70 & .... \\
 SS11 &  7.15 &3.36 & 77\\
 SS12 & 2.00 &0.88  & 3 \\
 SS13 & < 1.74 	& <0.82 & .... \\
 Stacked & & & 21
\enddata

\tablecomments{Col. (1): Object name. Col. (2): 3.3 \micron~PAH line flux. Col. (3). 3.3 \micron~PAH line luminosity.
Col. (4). Equivalent width of the 3.3 \micron~PAH emission line.
}
\end{deluxetable}

\begin{figure}
\epsscale{1.2}
\plotone{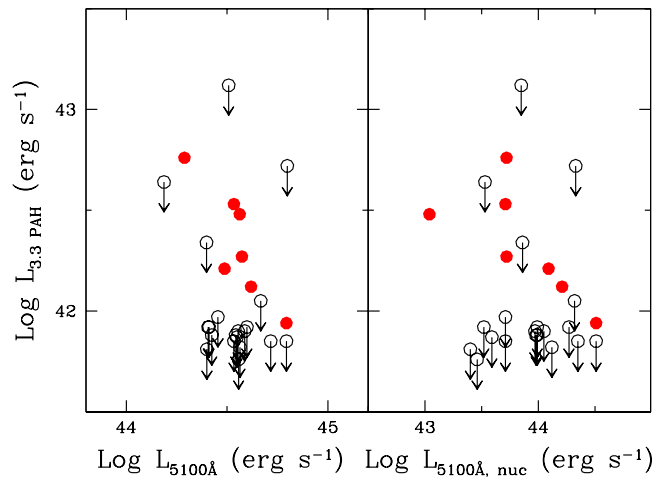}
\caption{Comparison of the 3.3\micron~PAH luminosity (\LPAH) with
the observed total luminosity at the rest-frame 5100 $\AA$ (L$_{\rm 5100\AA}$) ({\it left})
and AGN luminosity corrected for host galaxy starlight contribution ({\it right}). 
Filled circles represent the PAH detected galaxies while open circles represent
upper limits of \LPAH. Note that the monochromatic luminosity at 5100\AA~
includes the contribution from
starlight, which is substantial fraction of the total flux (74\% on average).
\label{fig3}}
\end{figure}

\section{Starburst-AGN Connection}
\label{results}

In this section, we investigate how starburst and AGN activity
is related. First, we compare AGN activity with the global star formation 
rate measured from the AKARI observations (\S~4.1).
Then, we investigate the correlation between nuclear starburst and
AGN activity based on the collected data from the literature (\S~4.2).

\subsection{Global PAH emission and AGN activity}
\label{Global}

To investigate the relation between global star formation and AGN activity,
we compare the luminosity of the 3.3 \micron~PAH emission (\LPAH) with various 
AGN properties, i.e., optical, X-ray, and bolometric luminosities.
First, in Figure 3 (left) we compare the \LPAH~with AGN continuum luminosity 
at 5100\AA~(L$_{\rm 5100\AA}$) measured from high S/N optical spectra 
obtained with the Low Resolution Imaging Spectrograph at the Keck Telescope \citep{woo06,bennert10}.
Although the optical luminosity range is rather limited, we find no
strong positive correlation. Instead, a weak negative correlation between \LPAH~
and the optical luminosity seems to be present among PAH-detected objects.
Assuming that the strengths of the 3.3 and 7.7\micron~PAH emissions are correlated,
this result is inconsistent with previous studies \citep[e.g.,][]{ntz07,lutz+08}, 
which reported a positive correlation between 7.7\micron~PAH luminosity 
and 5100\AA~luminosity. 
One potential bias is the contribution from host galaxy starlight to L$_{\rm 5100\AA}$ 
since for low luminosity AGN, such as the sample considered here, the amount
of starlight can be comparable to the AGN flux in the optical range \citep[e.g.,][]{woo06, bennert10}.

To correct for the host galaxy contribution, we measured the nuclear luminosity at 5100\AA~
based on the 2-dimensional AGN-host galaxy decomposition analysis using high-resolution 
images obtained with the Hubble Space Telescope (for details, see Treu et al. 2007; Bennert et al. 2010). 
The mean AGN-to-total flux ratio of our sample is 0.26 (see Table 1), 
indicating that the correction for the host galaxy contribution at 5100\AA~ is necessary.
After correcting for the host galaxy contribution, 
we compare the pure AGN optical luminosity at 5100\AA~ with the PAH luminosity 
as shown in Figure 3 (right).  
The nuclear luminosity (L$_{\rm 5100,nuc}$) becomes significantly smaller than the observed luminosity at 5100\AA. 
However, this correction does not significantly change the trend 
between \LPAH~and AGN continuum luminosity. 

These results are rather contradictory to the previous studies that claimed positive 
correlations between AGN activity and various starburst indicators, i.e., 
\LPAH~and far-IR luminosity \citep[e.g.][]{ntz07,Wt08,Oi10}. 
Note that there are many Type 1 AGN without PAH detections in the previous studies
as well as our observation.   
By including these upper limits,
the correlation between PAH and AGN luminosities can substantially weaken.
On the other hand, since the luminosity range of our sample is rather limited,
we cannot conclude whether or not there is a correlation between \LPAH~and L$_{\rm 5100\AA}$ in a larger luminosity range.

Instead of optical luminosities, we can use X-ray luminosity, which is
a good indicator of AGN activity since X-ray is dominantly radiated from
accretion disk in Type I AGN.
In Figure 4 (left), we compare \LPAH~with AGN X-ray luminosity integrated 
over the 0.5 to 8 keV range, 
measured from Chandra X-ray images \citep{szathmary11}.
Within the order of magnitude range of the X-ray luminosity,
we see no strong positive correlation.
We also used the bolometric luminosity of AGN determined from multi-wavelength SED fitting
analysis \citep[for details, see][]{szathmary11} to compare with \LPAH. As shown in Figure 4
(right), we see no strong correlation between \LPAH~ and the AGN bolometric luminosity
within the limited range of the observed luminosity.

\begin{figure}
\epsscale{1.2}
\plotone{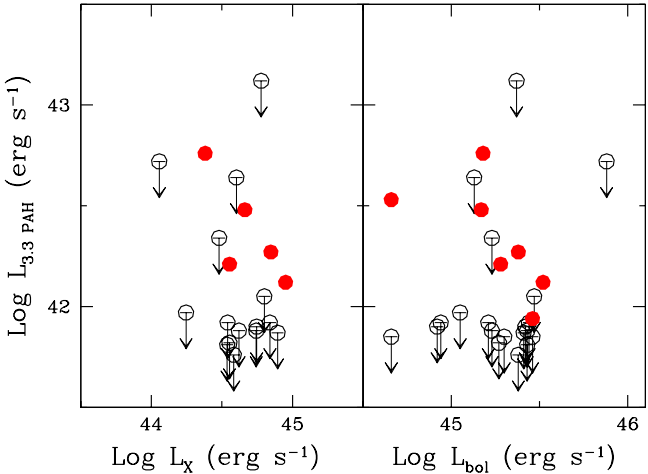}
\caption{The correlation of $L_{3.3 PAH}$ with X-ray luminosity ($L_{0.5-8 Kev}$)
measured from Chandra X-ray images. 
Symbols are same as in Figure \ref{fig3}. 
\label{fig5}}
\end{figure}

\begin{figure}
\epsscale{1.1}
\plotone{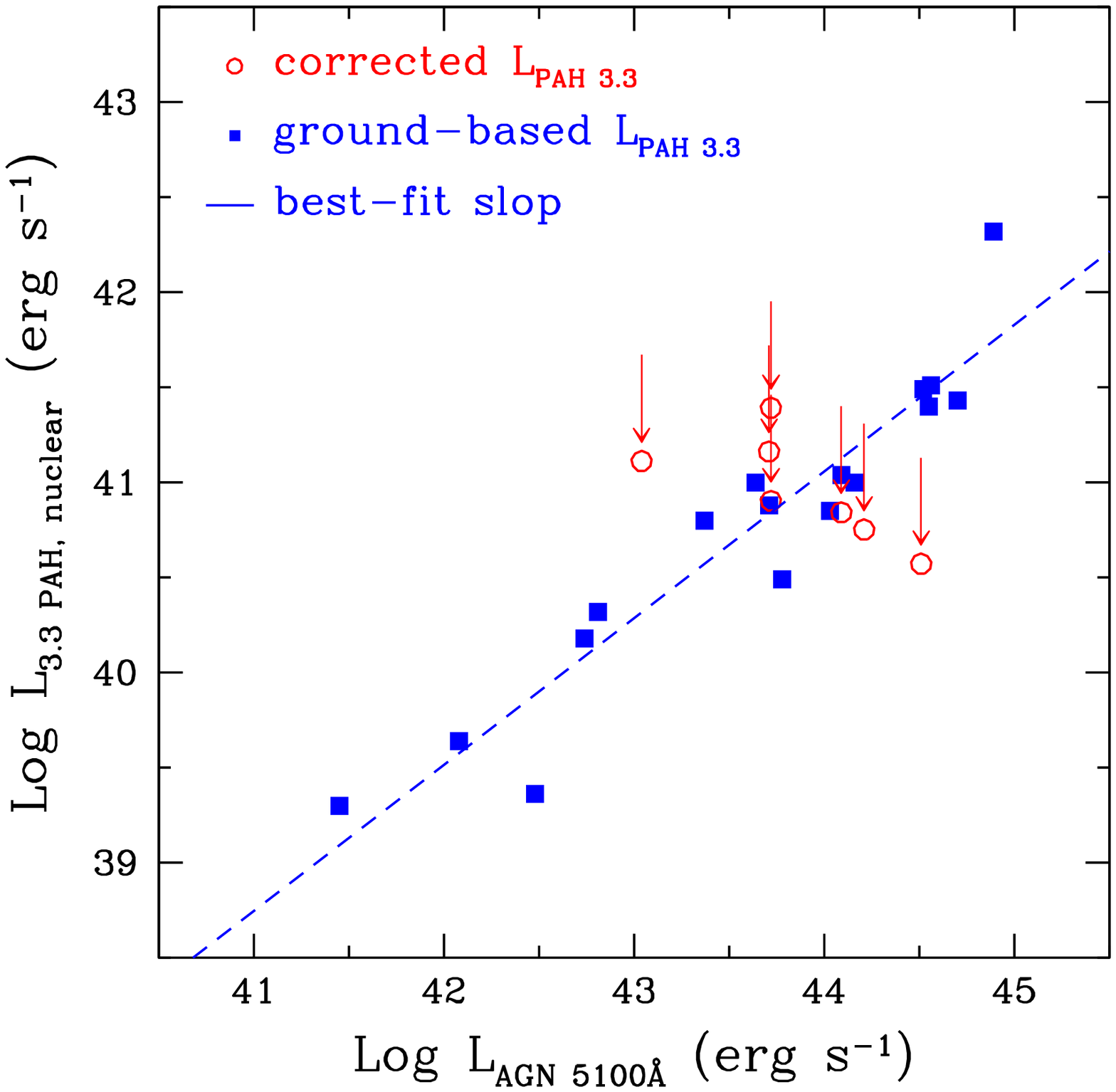}
\caption{The correlation of nuclear \LPAH~with AGN optical luminosity.
Filled squares represent \LPAH~ measurements from ground-based spectroscopy with 
a narrow-slit while open circles represent the estimated nuclear \LPAH~ 
of the AKARI sample, after dividing by a factor of
$\sim$23, which is the ratio between global and nuclear PAH luminosity of NGC 7469.
\label{fig8}}
\end{figure}

\begin{deluxetable*}{ccccccccccc}
\tablecaption{Nuclear 3.3\micron~PAH luminosity and AGN properties
\label{table3} }
\tablewidth{0pt}
\tablehead{ 
\colhead{Object}                   & 
\colhead{z}                   & 
\colhead{log L$_{\rm PAH 3.3}$}            &
\colhead{Ref.}                     &
\colhead{log M$_{\rm BH}$}         &
\colhead{Method}                   &
\colhead{Ref.}                     &
\colhead{log $\lambda$L$_{5100,nuc}$}  &
\colhead{Method}                   &    
\colhead{Ref.}                     \\
\colhead{} 
&\colhead{} 
&\colhead{(erg/s)}
&\colhead{}
&\colhead{(M$_{\odot}$)}
&\colhead{}
&\colhead{}
&\colhead{(erg/s)}
&\colhead{}                            
&\colhead{}                         \\
\colhead{(1)}                        &
\colhead{(2)}                        &
\colhead{(3)}                        &
\colhead{(4)}                        &
\colhead{(5)}                        &
\colhead{(6)}                        &
\colhead{(7)}                        &
\colhead{(8)}                        &
\colhead{(9)}                        &
\colhead{(10)}                        \\
}
\startdata
NGC 3227               & 0.0039 & 39.36 &  2 &  7.60 & reverberation       & 2         &  42.48  & bulge-disk decomposition  & 2         \\
NGC 4235               & 0.0008 & 39.64 &  1 &  7.71 & single-epoch virial & this work &  42.08  & from L$_{\rm H\alpha}$    & this work \\
NGC 4748 (MCG-2-33-34) & 0.0146 & 40.18 &  1 &  6.39 & reverberation       & 1         &  42.74  & from spectral fitting & 1 \\
NGC 5273               & 0.0036 & 39.30 &  4 &  6.85 & single-epoch virial & this work &  41.45  & from L$_{\rm H\alpha}$    & this work \\
NGC 5940               & 0.0339 & 40.80 &  1 &  7.95 & single-epoch virial & this work &  43.37  & from L$_{\rm H\alpha}$    & this work \\
Mrk 335                & 0.0258 & 40.49 &  1 &  7.13 & reverberation       & 2         &  43.78  & bulge-disk decompositoin & 2      \\
Mrk 509                & 0.0344 & 41.00 &  1 &  8.14 & reverberation       & 2         &  44.16  & bulge-disk decompositoin & 2      \\
Mrk 530                & 0.0295 & 40.88 &  1 &  8.06 & \msigma~relation    & 3         &  43.71  & from L$_{\rm bol.}$          & 3      \\
Mrk 618                & 0.0356 & 40.85 &  1 &  8.34 & M$_{\rm BH}$-L$_{\rm bulge}$ relation & 4     &  44.03  & from M$_{\rm BH}$ and Eq. 1     & 4      \\
Mrk 766 (NGC 4253)     & 0.0129 & 40.32 &  2 &  6.23 & reverberation       & 1          &  42.81  & from spectral fitting    & 1 \\
Mrk 817                & 0.0315 & 41.00 &  1 &  7.67 & reverberation       & 2          &  43.64  & bulge-disk decomposition & 2 \\
3C 120                 & 0.0330 & 41.04 &  1 &  7.72 & reverberation       & 2          &  44.09  & bulge-disk decomposition & 2 \\
PG 0157+001 (Mrk 1014) & 0.1631 & 42.32 &  3 &  8.31 & single-epoch virial & 5          & 44.89  & spectral fitting         & 5 \\
PG 1211+143            & 0.0809 & 41.43 &  3 &  8.14 & reverberation       & 2          &  44.70  & bulge-disk decomposition & 2 \\
PG 1411+442            & 0.0896 & 41.49 &  3 &  8.63 & reverberation       & 2          &  44.52  & bulge-disk decomposition & 2 \\
PG 1416-129            & 0.1289 & 41.40 &  3 &  8.83 & single-epoch virial & 5          &  44.55  & spectral fitting         & 5 \\
PG 1440+356            & 0.0791 & 41.51 &  3 &  7.80 & single-epoch virial & 6          &  44.54  & spectral fitting         & 6 \\
\enddata

\tablecomments{ 
Col. (1): Object name.
Col. (2): Redshift.
Col. (3): 3.3\micron~PAH luminosity 
Col. (4): reference for \LPAH. 
1. \citet{imanishi04}.
2. \citet{Rod03}
3. \citet{imanishi11}.
4. \citet{Oi10}.
5. \citet{imanishi10}.
Col. (5): Black hole mass
Col. (6): reference for black hole mass
1. Bentz et al. (2009b).
2. Peterson et al. (2004).
3. Nelson \& Whittle (1995).
4. Ryan et al. (2007).
3.
Col. (7): AGN monochromatic luminosity at 5100\AA, 
Col. (8): reference for L$_{5100}$.
1. Park et al. (2011).
2. Bentz et al. (2009a).
3. Woo \& Urry (2002).
4. Ryan et al. (2007).
5. Ho \& Kim (2009).
6. Shang et al. (2007).
}

\end{deluxetable*}

\subsection{Nuclear PAH emission and AGN activity}

In this section, we investigate how the nuclear 3.3\micron~PAH emission 
is related to AGN activity. We collected from the literature 
all 3.3\micron~PAH measurements of Type I AGN, for which AGN luminosity
and black hole mass can be compared. We found 3.3\micron~PAH measurements
for 20 local Seyfert 1 galaxies and PG QSOs based on the previous 
ground-based long-slit observations \citep{Rod03,imanishi04,Wt08,Oi10,imanishi11}.
These measurements based on the long-slit spectroscopy with a narrow slit 
represent the strength of the nuclear 3.3\micron~PAH emission.
Except for 3 AGN without necessary measurements of AGN properties in the literature, 
we determined black hole mass and AGN continuum luminosity at 5100\AA. 
Thus, we have a sample of 17 local Type I AGN as listed in Table 3.

For black hole masses, we adopted the reverberation mapping measurements
from Peterson et al. (2004) and Bentz et al. (2009b), after calibrating
with the most recent determination of the virial factor from \citet{woo10}.
When reverberation mapping results were not available, we used the single-epoch
virial method as used for the moderate-luminosity AGN sample.
In practice we use the following equation to estimate black hole masses (Park et al. 2011):
\begin{equation}
M_{\rm BH} = 10^{7.602}{\rm M}_{\odot} \left(
{\sigma_{{\rm H}\beta}({\rm rms}) \over 1000~ {\rm km~s}^{-1} } \right)^{2} \left( {\lambda
L_{5100,nuc} \over 10^{44}~ {\rm erg~s}^{-1}} \right)^{0.518}~,
\label{eq:MBH_sigmaRMS}
\end{equation}
where $\sigma_{\rm H\beta}$ is the line dispersion of the \hb~line and L$_{5100, nuc}$
is the AGN continuum luminosity at 5100\AA~corrected for the host galaxy starlight contribution.  
For NGC 4235, NGC 5273, and NGC5940 with no line width measurements in the literature, 
we fit the broad \ha~ emission line and measured the line width and luminosity 
using the spectra from the Sloan Digital Sky Survey.
For two objects (Mrk 530 and Mrk 618) without any available spectra,
we used the black hole mass -galaxy property relations, i.e., \msigma and
M$_{\rm BH}$-L$_{\rm bulge}$ relations \citep{gultekin+09} to calculate black hole masses
(see Table 3 for details).

In the case of AGN power, we collect AGN continuum luminosity
at 5100\AA~as a proxy for the bolometric luminosity of AGN.
For high luminosity QSOs, i.e., PG QSOs, it is not necessary to correct
for the host galaxy starlight contribution. In contrast, for low luminosity AGN,
L$_{\rm 5100\AA}$ should be corrected for the host galaxy contamination.
In general we obtained L$_{\rm 5100\AA, nuc}$ measurements from previous studies,
which were based on either bulge-disk decomposition analysis using HST images or 
spectral fitting analysis using high quality spectra. For 3 objects, namely NGC 4235, NGC5273,
and NGC 5940, we determined AGN continuum luminosity from \ha~line luminosity 
using the tight correlation between L$_{5100 \AA,nuc}$ and L$_{\rm H\alpha}$ 
\citep[see Eq. 1 and 2 in][]{greene&ho05}.

In contrast to the ground-based observations, \LPAH~ measured from AKARI observations represents the global starburst as discussed in \S~3.1.
Thus, in order to include the PAH emission measurements from AKARI
observations, it was necessary to calibrate them for representing nuclear
starburst. The relation between nuclear and global starburst for Type I AGN
has not been studied in detail \citep{imanishi10}. Although the 
nuclear-to-global starburst ratio can vary for individual objects,
we used the case of NGC 7469 to estimate the difference.
As discussed in \S~3.1, NGC 7469 was observed with a ground-based spectrograph
with a narrow-slit and the AKARI IRC, respectively. 
The flux ratio between nuclear \LPAH~ and the upper limit of the global \LPAH~is $\sim$0.04.
Thus we took this ratio as a calibration factor and scale down the AKARI
measurements by a factor of 0.04.

In Figure 5, we present the correlation between
the nuclear 3.3\micron~PAH luminosity and AGN optical luminosity at 5100\AA~
for the combined sample.
The Type I AGN from ground-based observations clearly show a correlation,
suggesting the nuclear starburst-AGN connection. The \LPAH~of
the moderate-luminosity AGN at z$\sim$0.4, which are calibrated for representing
the nuclear PAH emission, are consistent with the trend found by
nuclear \LPAH~measurements although they are more scattered probably 
due to the uncertainty of the calibration. 

Using the nuclear \LPAH~measurements from ground-based spectroscopy,
we fit the relation between \LPAH~and AGN luminosity.
Since most of the \LPAH~measurements in the literature do not have error estimates,
we fit the best-slope without accounting for the measurement uncertainties.
The best-fit relation is:
\begin{equation}
\log ({\rm L_{\rm PAH 3.3} \over 10^{41}~ {\rm erg s^{-1}}}) = 0.06\pm0.01 + (0.77\pm0.01) 
\times \log ({\rm L_{\rm AGN~5100\AA} \over 10^{44}~ {\rm erg s^{-1}}}).
\label{eq:msigma}
\end{equation}
The best-fit slope of $\sim$0.8 indicates that the correlation is not linear.
This implies that for higher luminosity AGN, the nuclear starburst-to-AGN luminosity 
ratio is smaller, albeit with the small sample size.

In Figure 6, we compare nuclear \LPAH~with the Eddington ratio and black hole mass
to test whether the accretion rate and black hole mass correlates 
with starburst activity. 
We estimate the Eddington ratio by multiplying a factor
of 9 to the AGN optical luminosity.
In the case of the Eddington ratio, there is a general trend that higher
Eddington ratio objects show higher nuclear starburst activity although
there is a large scatter.
Black hole mass shows a similar trend with \LPAH~in that the host galaxies
of higher mass AGN show higher nuclear starburst luminosity. This can
be interpreted as the trend that higher mass AGN in the sample are
generally higher luminosity AGN, i.e., PG QSOs. Thus, the relation
between black hole mass and PAH luminosity reflects the correlation
between AGN power and \LPAH.

\begin{figure}
\epsscale{1.2}
\plotone{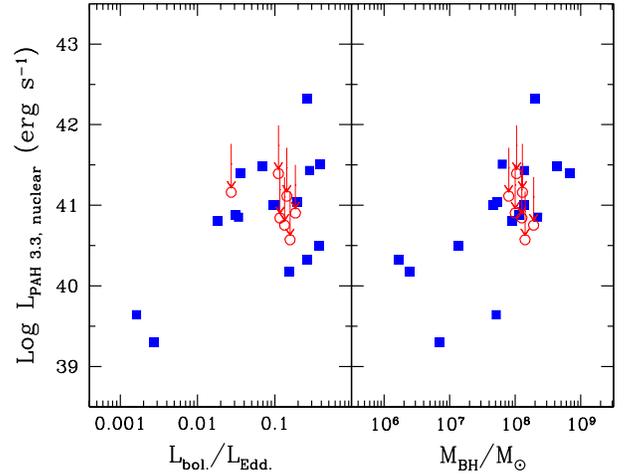}
\caption{Correlation of nuclear \LPAH~with AGN optical luminosity.
Filled squares represent \LPAH~measurements from ground-based spectroscopy with 
a narrow-slit while open circles represent the estimated nuclear \LPAH~ 
of AKARI sample, after dividing by a factor of
$\sim$23, which is the ratio between global and nuclear PAH luminosity 
of NGC 7469.
\label{fig6}}
\end{figure}

\section{Discussion and Conclusions}

To investigate the starburst-AGN connection, we compared
the global and nuclear 3.3 \micron~PAH emission with AGN
luminosity using a sample of Type I AGN.
The global 3.3 \micron~PAH emission luminosities are measured
for moderate-luminosity AGN at z$\sim$0.4, 
based on our new AKARI slit-less spectroscopy, which covers the 
entire host galaxies. 
We find no strong correlation between 3.3 \micron~PAH and AGN optical 
luminosities within the limited AGN luminosity range 10$^{44}$ - 10$^{45}$ \ergs.
This result may imply that the global star formation is not strongly 
related with AGN activity although we cannot rule out a weak correlation 
with a considerable scatter. We note that since the AGN luminosity range
is very limited, whether the global star formation rate
correlates with AGN activity is inconclusive.

In contrast, when we compare AGN luminosity of the local Seyfert 1 galaxies
and PG QSOs with their nuclear 3.3 \micron~PAH emission luminosity,
we detect a strong correlation,
suggesting that that AGN activity is related to the nuclear starburst. 
These results are consistent with the findings of previous studies on the starburst-AGN
connection \citep{Im03, imanishi04, ntz07, lutz+08, Oi10}.

The slope of the \LPAH~correlation with AGN luminosity in Figure 5 shows  
that the \LPAH~to AGN luminosity ratio decreases toward higher AGN luminosity.
At face value, it implies that the nuclear starburst activity 
is slightly suppressed in the host galaxies of high luminosity AGN (for different predictions
on the starburst-to-accretion ratios, see e.g., Kawakatu \& Wada 2008; Ballantyne 2008).
A similar trend was noted by \cite{lutz+08} in comparing 7.7 \micron~PAH emission
and AGN luminosities. 
In their study, high luminosity AGN with L$_{\rm 5100\AA}$ $>$ 10$^{46}$ \ergs~
show a much lower starburst-to-AGN ratio \citep[see Figure 6 in][]{lutz+08}, 
implying that the black hole growth rate normalized by the starburst rate is a factor of
10 higher in their high redshift QSOs than in the local AGN.
In our local sample the decrease of the starburst-to-AGN ratio with increasing
AGN luminosity is not as dramatic as in high redshift QSOs, but if confirmed, 
it implies that for low-luminosity AGN 
the growth of galaxy centers is faster than black hole growth in the present-day universe.
However, it is not clear whether black hole growth is slower or faster than the total host
galaxy growth since only nuclear starbursts are compared in our analysis.

We note that there are many AGN with no PAH emission detection. 
Including the upper limits of
these objects significantly weakens the correlation between \LPAH~and AGN luminosity.
Although the true PAH luminosities of these objects may be close to the upper limits 
within a factor of few, a detailed investigation using a large sample with lower
flux limits is necessary to probe the slope and scatter of the correlation.  

\acknowledgements 
This work has been supported by the Basic Science Research Program through the National Research Foundation of Korea (NRF) funded by the Ministry of Education, Science and Technology (MEST), No. 2010-0021558. 
M.I. acknowledges support by Grants-in-Aid for Scientific Research no. 22012006. 
This research is based on observations with AKARI, a JAXA project with the participation of ESA.


\end{document}